\documentclass[]{gPHT2e}

\begin{document}



\title{Monte Carlo study of the XY-model on Sierpinski gasket}

\author{Bo\v zidar Mitrovi\' c$^{\ast}$\thanks{$^\ast$Corresponding author. Email: bmitrovic@brocku.ca
\vspace{6pt}} and Shyamal K.~Bose\\\vspace{6pt}  {\em{Department of Physics, Brock University, St.~Catharines, 
Ontario, Canada L2S 3A1}}\\\vspace{6pt}\received{} }

\maketitle

\begin{abstract}
We have performed a Monte Carlo study of the classical XY-model on  two-dimensional Sierpinski  
gaskets of several cluster sizes. From the dependence of the helicity modulus on the cluster size we
conclude that there is no phase transition in this system at a finite temperature. This is in agreement  
with previous findings for the harmonic approximation to the XY-model on Sierpinski gasket and is 
analogous to the absence of finite temperature phase transition for the Ising model on fractals with 
a finite order of ramification.\bigskip

\bigskip

\end{abstract}

\section{Introduction}

In a series of papers Gefen, Mandelbrot and Aharony \cite{gab80,gab83,gasb84,gab84} examined 
the possibility of phase transitions for discrete-symmetry spin models (e.g.~the Ising model) on  
self-similar fractal structures. By applying renormalization-group techniques they found that the critical 
properties of such systems vary with several topological characteristics: the (noninteger) fractal dimensionality
 $D$, the order of ramification $R$ and lacunarity \cite{gab80}. They established that a lower  
critical dimensionality is not defined and found that the Ising model on systems with given $D$ have transition 
temperature $T_{c}=$0 if the minimum order of ramification $R_{min}$ is finite, and $T_{c}>$0 if 
$R_{min}$ is infinite. The order of ramification $R$ at a point $P$ of a structure is defined as the number 
of bonds which one must cut in order to isolate an arbitrarily large bounded set of points connected to $P$. 
For two-dimensional Sierpinski gasket (SG) $R_{min}$=3 and $R_{max}$=4 \cite{gab80}. 
Subsequently, Monceau and Hsiao \cite{mh04} did a Monte Carlo study of the Ising model on fractals with  
infinite $R_{min}$ which had the same fractal dimension 1$<D<$3 but different structure. They found direct 
evidence of so-called weak universality in such systems, i.e.~ the critical exponents depend not only on 
$D$, the symmetry of the order parameter and the range of interactions but also on topological features of 
the fractal, in particular on lacunarity which measures the deviation from translational symmetry. 

The question remained as to what happens in the case of $n$-component spin models, with $n\geq$2, 
on fractal structures. In \cite{gasb84,gab84} a correspondence between pure resistor network connecting  
sites of a given lattice and the low-temperature properties of the spins with continuous symmetry ($n\geq$2)  
on the same lattice \cite{stinc79} was used to conclude that there is no long-range order at 
any finite temperature if the fractal dimension $D<$2 even in the case of infinite order of ramification. 
For two-dimensional SG $D=\ln$3/$\ln$2=1.585.
Subsequently, in their study of the effect of phase fluctuations of the superconducting order parameter 
in two-dimensional Sierpinski gasket wire networks \cite{gg86} Vallat, Korshunov and Beck 
\cite{vkb91} examined the XY-model on such a lattice in the harmonic approximation. They found 
that the energy of a vortex is always finite on the two-dimensional Sierpinski gasket implying that there is no 
Berezinskii-Kosterlitz-Thouless transition \cite{stran88} associated with vortex-antivortex unbinding as 
free vortices are always present.

Here we present a Monte Carlo study of the XY-model on two-dimensional Sierpinski gaskets described by 
the Hamiltonian
\begin{equation}
\label{ham}
 H=-J\sum_{\langle i,j\rangle}\cos(\theta_{i}-\theta_{j})\>,
\end{equation}
where 0$\leq \theta_{i}<$2$\pi$ is the angle variable on site $i$, $\langle i,j\rangle$ denotes the 
nearest neighbors and $J>$0 is the coupling constant. 
In the case of XY-model on periodic and quasiperiodic lattices \cite{rbm98} the most 
important excitations at low temperatures are the spin-waves and the harmonic approximation to (\ref{ham}) 
used in \cite{vkb91} is quite good for the purpose of studying the low-temperature behavior. 
We compute the heat capacity, the helicity modulus and susceptibility 
for Sierpinski gaskets of several sizes. The size dependence of the helicity modulus and its sensitivity 
to the boundary conditions clearly indicate that in the thermodynamic limit there is no phase transition at 
finite temperature.     

The rest of the paper is organized as follows. In Section~\ref{numerics} we describe numerical 
procedure used in calculations and present our numerical results. In Section~\ref{summary} we 
summarize our conclusions. 

\section{The numerical procedure and the results}\label{numerics}

The procedure which we used to generate two-dimensional Sierpinski gaskets is illustrated in Figure 1, which 
shows the transition from the zeroth-order Sierpinski gasket (the equilateral triangle of side $a$) to the 
first order gasket. Our basic unit was the fourth order Sierpinski gasket with 123 sites (a gasket of order 
$m$ has $N$=3(3$^{m}$+1)/2 sites) and the list of the nearest neighbors of each site. The higher order  
Sierpinski gaskets were created as subsequent generations of the fourth order gasket and we considered gaskets  
up to order $m$=7 with 3282 sites. In going from the gasket of order $m$ to the gasket of order $m$+1 the 
list of the nearest neighbors for sites close to the three corners of each of the three $m$th order gaskets 
had to be modified.

\begin{figure}
\begin{center}
\resizebox*{8cm}{!}{\includegraphics{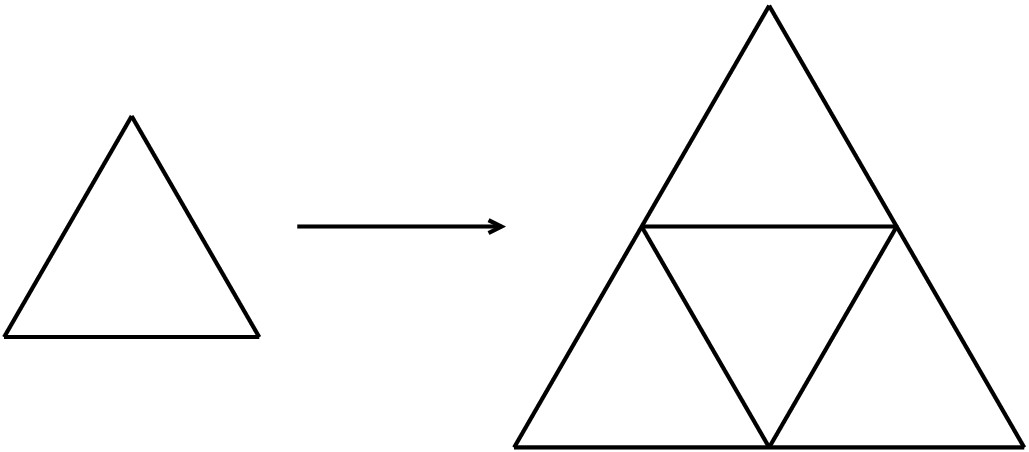}}%
\caption{\label{fig1} The first step in creating Sierpinski gaskets.}
\label{fig1}
\end{center}
\end{figure}

To study the statistical mechanics of the Hamiltonian (\ref{ham}) on Sierpinski gaskets we have used 
Monte Carlo (MC) simulations based on Metropolis algorithm \cite{metro53}. For a gasket of given order 
the simulation would start at a low temperature with all phases aligned. The first $k$ MC steps per site (sps) 
were thrown away, followed by seven MC links of $k$ MC sps each. The values of $k$ which we used were 
120000, 360000, and 600000, depending on the system size and statistical errors. At each temperature the 
range over which each angle $\theta_{i}$ was allowed to vary \cite{bh88} was adjusted to ensure an MC 
acceptance rate of about 50\%. The errors were calculated by breaking up each link into six blocks of 
$k$/6 sps, then calculating the average values for each of 42 blocks and finally taking the standard deviation 
$\sigma$ of these 42 average values as an estimate of the error. The final configuration of the angles 
$\{\theta_{1},\dots,\theta_{N}\}$ at a given temperature was used as a starting configuration for the 
next higher temperature. Two types of boundary conditions were considered: closed, where the three corners of 
an $m$th order gasket were considered to be coupled to each other
(then each site has four nearest neighbors) and open, where
the three corners are uncoupled to each other. 

In Figure 2 we show the results for the heat capacity per site calculated from the fluctuation theorem 
\begin{equation}
\label{cv}
C=\frac{1}{N}\frac{\langle H^{2}\rangle - \langle H\rangle^{2}}{k_{B}T^{2}}\>,
\end{equation}

\begin{figure}
\begin{center}
\resizebox*{10cm}{!}{\includegraphics{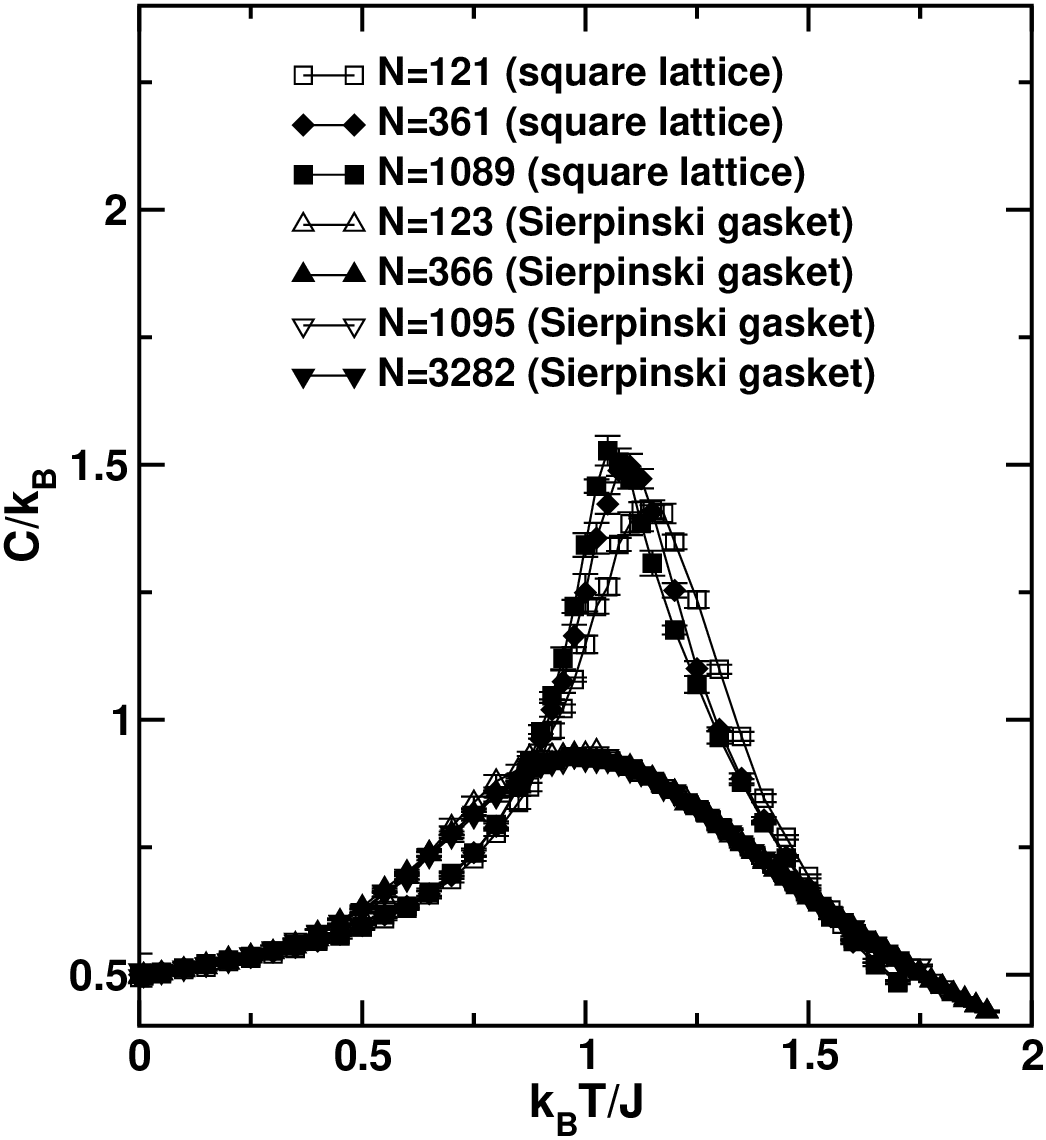}}%
\caption{\label{fig2} Calculated heat capacity per site as a function of temperature.}
\label{fig2}
\end{center}
\end{figure}

\noindent where $k_{B}$ is the Boltzmann constant, $T$ is the absolute temperature and $\langle\cdots\rangle$  
denotes the MC average. The results for the four Sierpinski gaskets were obtained with the closed boundary  
conditions. In the case of heat capacity the results do not depend on boundary conditions as 
illustrated in Figure 3. For comparison, we have included in Figure 2 the results obtained 
for the Hamiltonian (\ref{ham}) on three square lattices with the periodic boundary conditions. 
The sizes of the square lattices were chosen so that the number of sites is similar to the number of sites for 
the three smaller Sierpinski gaskets. As there is no long-range order in two dimensions the heat capacity per
site of the square lattices saturates with increasing system size \cite{tc79}. In the case of Sierpinski 
gaskets there is virtually no size dependence of the heat capacity per site. This, however, does not mean 
that the model (\ref{ham}) on two-dimensional SG also leads to  
the Berezinskii-Kosterlitz-Thouless (BKT) transition resulting from vortex-antivortex unbinding at 
transition temperature $T_{c}$. The peak in the heat capacity of the square lattices is above the BKT 
transition temperature where the heat capacity has an unobservable essential singularity \cite{bn79}. The peak  
is caused by unbinding of vortex clusters \cite{tc79} with increasing temperature above $T_{c}$. In the 
case of Sierpinski gaskets the broad peak in $C$ could result from the average energy per site 
$\langle E\rangle$ changing monotonically from the values near -2$J$ (each site has four nearest 
neighbors) at low temperatures to near zero at high temperatures (disordered paramagnetic phase). 

\begin{figure}
\begin{center}
\resizebox*{10cm}{!}{\includegraphics{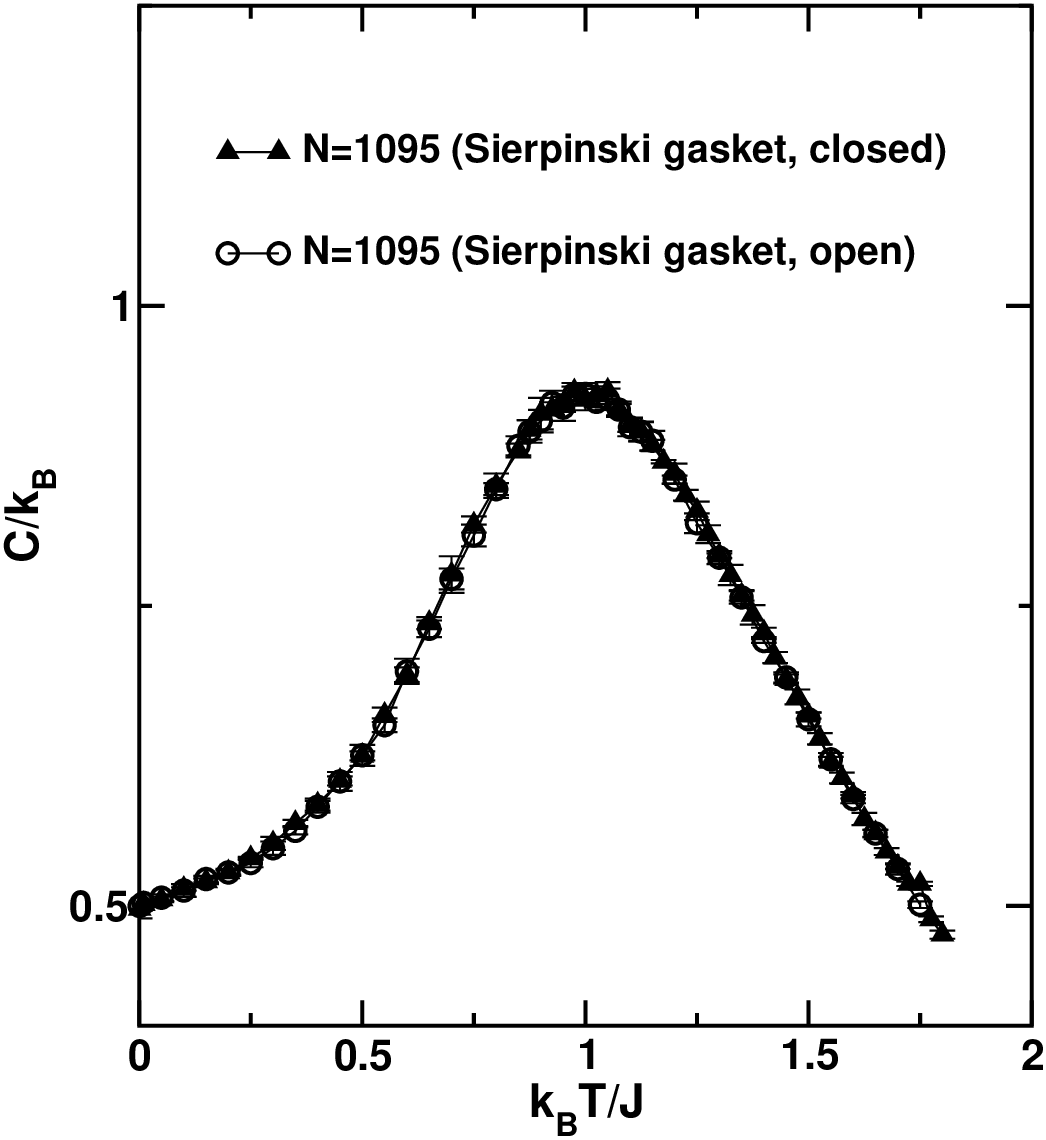}}%
\caption{\label{fig3} The heat capacity per site as a function of temperature for the sixth 
order SG calculated with closed boundary condition (full triangles) and with 
the open boundary condition (open triangles).} 
\label{fig3}
\end{center}
\end{figure} 

A much better indicator of BKT transition is temperature dependence of the
helicity modulus. It measures the stiffness of the angles $\{\theta_{i}\}$ with respect to a twist 
at the boundary of the system. When the Hamiltonian (\ref{ham}) is used to model a set of identical
superconducting grains coupled via Josephson tunneling, in which case $\theta_{i}$ is the phase of the
superconducting order parameter on grain $i$, the effect of applied magnetic field described by the vector
potential ${\bf A}$ is included via shift in the phase difference $\theta_{i}-\theta_{j}$ by 
\begin{equation}
\label{shift}
A_{ij}=\frac{2\pi}{\Phi_{0}}\int_{{\bf r}_{i}}^{{\bf r}_{j}}d{\bf r}\cdot{\bf A}\>,
\end{equation}
where the line integral is taken along the line joining sites $i$ and $j$ and $\Phi_{0}=hc/2e$ is the flux 
quantum \cite{es83}. For a {\em uniform} vector potential ${\bf A}$ one has 
$A_{ij}=2\pi{\bf A}\cdot({\bf r}_{j}-{\bf r}_{i})/\Phi_{0}$. Then, the helicity modulus can be defined 
\cite{es83} as
\begin{equation}
\label{hel}
\gamma=\left(\frac{\partial^{2}F}{\partial A^{2}}\right)_{A=0}
\end{equation}
where $F$ is the Helmholtz free energy. Using $\partial^{2}F/\partial A^{2}=
\langle\partial^{2}H/\partial A^{2}\rangle-\langle(\partial H/\partial A)^{2}\rangle/(k_{B}T)+
(\langle\partial H/\partial A\rangle)^{2}/(k_{B}T)$ one finds an expression for the helicity modulus which is 
analogous to (\ref{cv}). For the Sierpinski gaskets ${\bf A}$ was taken to be parallel to one of the sides of 
the triangles (Figure 1) and for the square latices ${\bf A}$ was taken to be parallel to one of the sides of 
the squares. Our MC results for $\gamma$ obtained with closed boundary condition for the Sierpinski gaskets 
and with periodic boundary conditions for the square lattices are shown in Figure 4. One expects a finite 
$\gamma$ at low temperature when the angles are in an ordered configuration and zero stiffness in 
high temperature paramagnetic phase. Nelson and Kosterlitz \cite{nk77} predicted a discontinuous jump 
in $\gamma$ at the BKT transition temperature $T_{c}$ with a universal value $\gamma(T_{c})/T_{c}$=2/$\pi$,  
and the straight line in Figure 4 gives the value of the jump at different temperatures. For a finite system 
the jump in $\gamma$ is replaced by continuous decrease with increasing $T$ which becomes steeper near the
(putative) $T_{c}$ with increasing system size, as indicated by the results for the square lattices in  
Figure 4. The two important aspects of our results are: (1) For all lattices considered a rapid downturn in  
$\gamma$ starts near the universal 2/$\pi$-line. (2) While the low temperature values of $\gamma$ for the  
square lattices do not depend on the system size, for the Sierpinski gaskets they clearly decrease with 
increasing system size and the onset of the downturn in $\gamma$, which is in the vicinity of the 
universal 2/$\pi$-line, moves to the lower temperatures. This suggests that in the thermodynamic limit 
(i.e.~ when the order of the SG $m\rightarrow\infty$) the angle stiffness $\gamma$ would 
be zero at any finite temperature implying no BKT transition at a finite temperature.    
\begin{figure}
\begin{center}
\resizebox*{10cm}{!}{\includegraphics{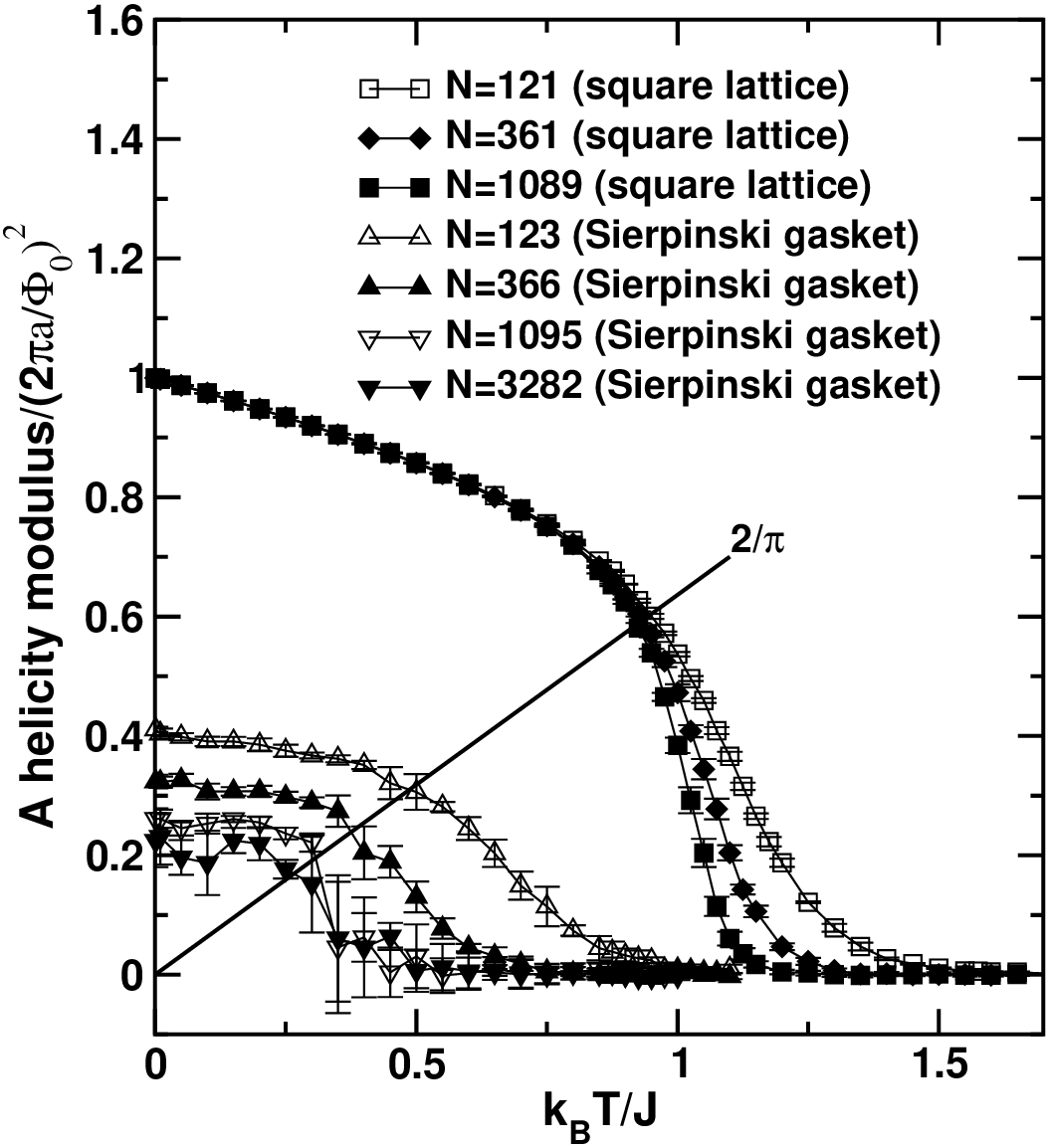}}%
\caption{\label{fig4}  The helicity modulus $\gamma$ as a function of temperature for Sierpinski gaskets 
and square lattices of different sizes. The straight line gives the size of Kosterlitz-Thouless discontinuous 
jump in $\gamma$ at various temperatures.} 
\label{fig4}
\end{center}
\end{figure}

We have examined the effect of boundary conditions (closed vs. open boundary conditions) and the 
results for the sixth order SG are shown in Figure 5. While the boundary conditions had 
no effect on the specific heat (see Figure 3) their effect on the helicity modulus is dramatic. The open 
boundary conditions give zero helicity modulus at all temperatures within the error bars, which are much 
larger when the three corners of the sixth order SG are not coupled to each other. Such a 
dramatic change in $\gamma$ implies that the closed boundary condition introduces significant additional 
correlations in two-dimensional SG fractal structure. Results obtained with the open boundary 
condition reinforce our conclusion about the absence of phase transition at any finite $T$. 
\begin{figure}
\begin{center}
\resizebox*{10cm}{!}{\includegraphics{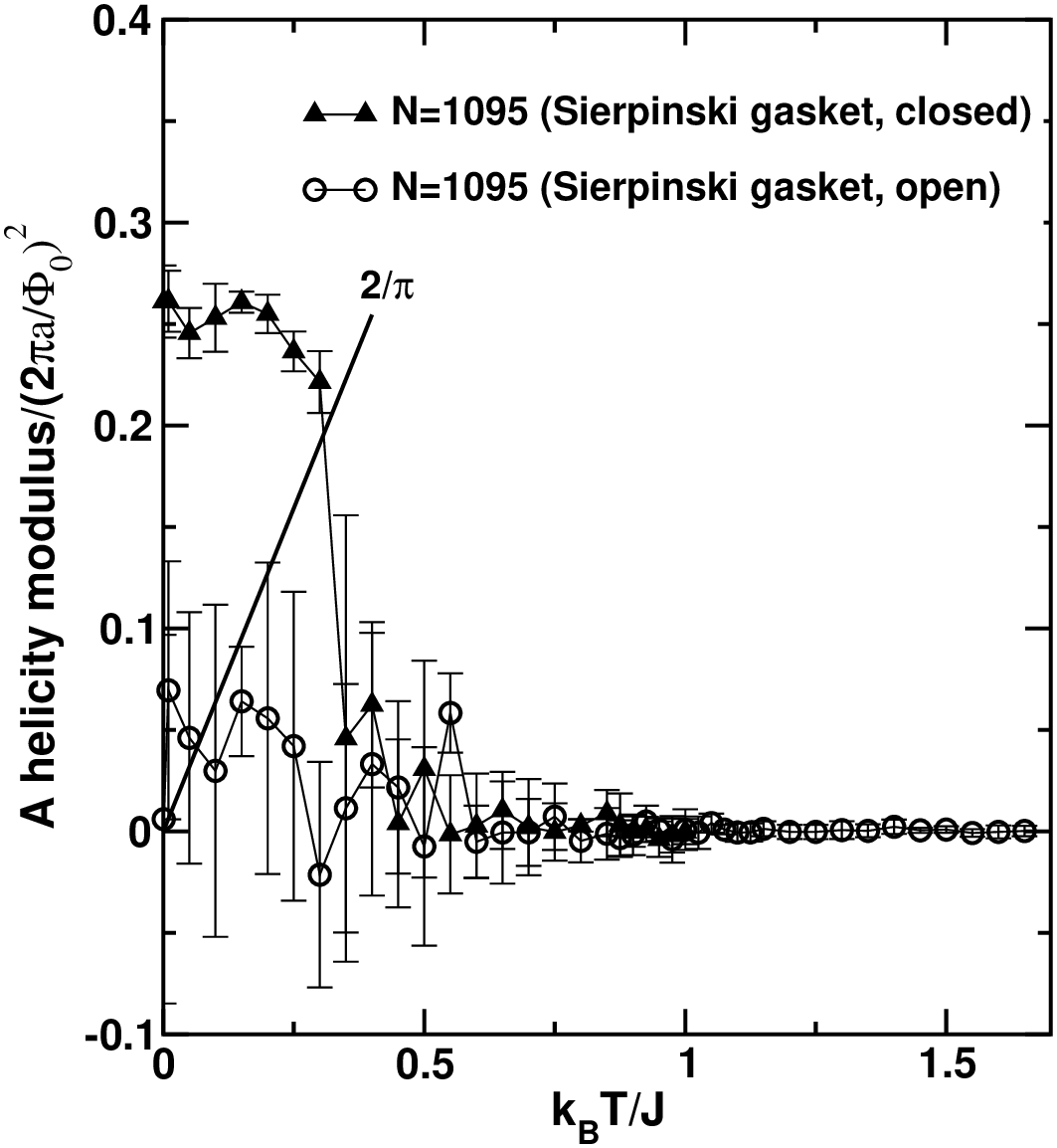}}%
\caption{\label{fig5}  The effect of boundary conditions on the temperature dependence of the 
the helicity modulus for SG of order $m$=6.}
\label{fig5}
\end{center}
\end{figure}

Finally, for completeness, in Figure 6 we show our results for the linear susceptibility per site computed from 
\begin{equation}
\label{susc}
\chi=\frac{\langle M^{2}\rangle-\langle M\rangle^{2}}{N^{2}k_{B}T}\>,
\end{equation}
where $M$ is the magnetization of the system when Hamiltonian (\ref{ham}) is used to describe magnetic systems. 
For BKT transition Kosterlitz predicted \cite{k74} that above $T_{c}$ the susceptibility diverges as 
$\chi\sim\exp[(2-\eta)b(T/T_{c}-1)^{-\nu}]$, with $\eta$=0.25, $b\approx$1.5 and $\nu$=0.5, and is 
infinite below $T_{c}$. For finite systems one gets finite peaks in $\chi$ above $T_{c}$ which increase in  
height and move to lower temperatures with increasing system size, 
as illustrated by our results for three square lattices in Figure 6. This trend is analogous to what 
one finds for (\ref{ham}) in three dimensions where there {\em is} long-range order below the transition 
temperature. Therefore the simulation results for susceptibility are not useful in diagnosing BKT transition 
unless one has the results for very large cluster size and attempts to fit them according to prediction by 
Kosterlitz \cite{rbm98}. We note that the susceptibilities for Sierpinski gaskets obtained with closed 
boundary condition are 
considerably higher than those for the square lattices of comparable sizes. Also, the error bars are 
considerably larger than what was obtained for the square lattices. The effect of boundary conditions 
on linear susceptibility for Sierpinski gaskets is shown in Figure 7. The main difference is that open 
boundary condition leads to much larger error bars at low temperatures which is consistent with our 
earlier observation that the closed boundary condition introduces additional correlations in the system. 
\begin{figure}
\begin{center}
\resizebox*{10cm}{!}{\includegraphics{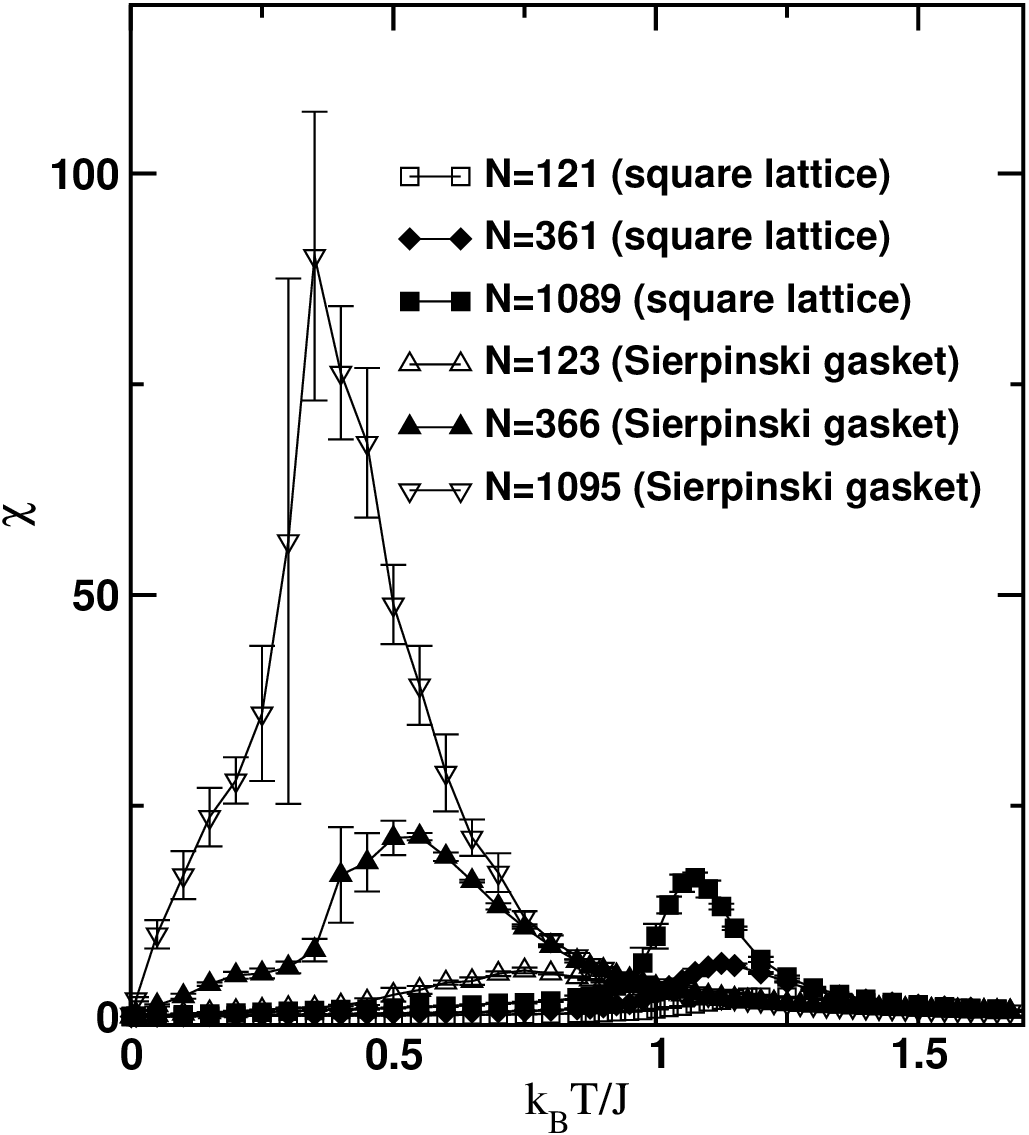}}%
\caption{\label{fig6} The linear susceptibility for Sierpinski gaskets and square lattices.} 
\label{fig6}
\end{center}
\end{figure}
\begin{figure}
\begin{center}
\resizebox*{10cm}{!}{\includegraphics{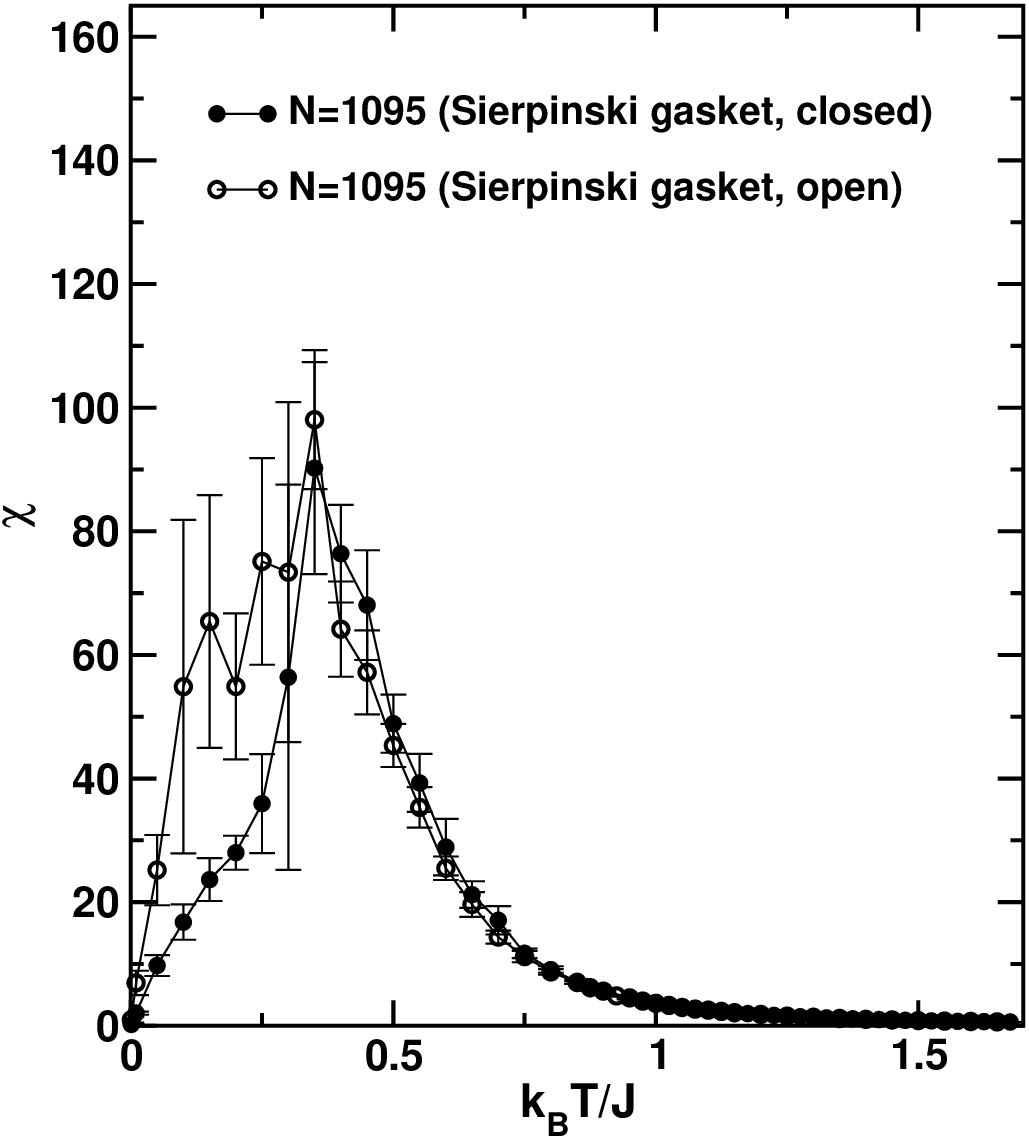}}%
\caption{\label{fig7} The effect of boundary conditions on the temperature dependence of the
the susceptibility for the Sierpinski gasket of order $m$=6. }%
\label{fig7}
\end{center}
\end{figure}

\section{Summary}\label{summary}

Our Monte Carlo results for the specific heat for the XY-model on two-dimensional Sierpinski gaskets show
that there is no long range order at any finite temperature as the specific heat is virtually independent of
the system size. Moreover, the cluster-size dependence of the low temperature helicity modulus clearly
indicates that there is no Berezinski-Kosterlitz-Thouless transition in this planar fractal structure in
the thermodynamic limit. The low temperature values of the helicity modulus $\gamma$ decrease
with increasing system size while they are size-independent for square lattices where the BKT
transition does take place. Moreover, in the case of Sierpinski gaskets
the onset of the downturn in helicity modulus near the universal 2/$\pi$-line moves to the lower
temperatures with increasing size of the clusters while for the square lattices only the rate of decline in
$\gamma$ beyond the 2/$\pi$-line increases with the system size approaching a discontinuous jump in
the thermodynamic limit. Our results are consistent with previous findings of Gefen et al. \cite{gasb84} who 
suggested, based on their exact renormalization group treatment of the resistor network on 
Sierpinski gasket, that spin systems with continuous symmetry on this structure do not display a finite 
temperature phase transition. Also, our findings are consistent with the results of Vallat et al.   
\cite{vkb91} obtained for the quadratic approximation to the XY-model on two-dimensional Sierpinski gasket, 
who found that the vortex energy is finite and hence free vortices are present at any temperature.  

\subsection{Acknowledgements}

This work was supported in part by the Natural Sciences and Engineering Research Council of Canada.

\section*{Note added in proof} After this manuscript was accepted for publication it came to our attention that the possibility of phase transitions on lattices of effectively nonintegral 
dimensionality was first examined by Dhar (D.~Dhar, {\em Lattices of effectively nonintegral dimensionality}, J.~Math.~Phys.~18 (1977),pp.~577--585). In the case of the classical XY-model on 
the truncated tetrahedron lattice (the effective dimensionality 2$\log$2/$\log$5) no phase transition at any
finite temperature was obtained. 


\begin{thebibliography}{17}
\markboth{Taylor \& Francis and I.T. Consultant}{Phase Transitions}
\bibitem[1]{gab80}
Y.~Gefen, B.B.~Mandelbrot, and A.~Aharony, {\em Critical Phenomena on Fractal Lattices}, Phys.~Rev.~Lett.~45 (1980), pp.~855--858.

\bibitem[2]{gab83}
Y.~Gefen, A.~Aharony, and B.B.~Mandelbrot, {\em Phase transitions on fractals: I.~Quasi-linear lattices}, J.~Phys.~A: Math.~Gen.~16 (1983), pp.~1267--1278.

\bibitem[3]{gasb84}
Y.~Gefen, A.~Aharony, Y.~Shapir, and B.B.~Mandelbrot, {\em Phase transitions on fractals: II.~Sierpinski 
gaskets}, J.~Phys.~A: Math.~Gen.~17 (1984), pp.~435--444.

\bibitem[4]{gab84}
Y.~Gefen, A.~Aharony, and B.B.~Mandelbrot, {\em Phase transitions on fractals: III.~Infinitely ramified 
lattices}, J.~Phys.~A: Math.~Gen.~17 (1984), pp.~1277--1289. 

\bibitem[5]{mh04}
P.~Monceau and P.-Y.~Hsiao, {\em Direct evidence for weak universality on fractal structures}, Physica A 331 
(2004), pp. 1--9.

\bibitem[6]{stinc79}
R.B. Stinchcombe, {\em Position space treatment of diluted classical Heisenberg ferromagnet}, J.~Phys.~C: Solid 
State Phys.~12 (1979), pp.~2625--2636.

\bibitem[7]{gg86}
J.M~Gordon, A.M.~Goldman, J.~Maps, D.~Costello, R.~Tiberio, and B.~Whitehead, {\em Superconducting-Normal 
Phase Boundary of a Fractal Network in a Magnetic Field}, Phys.~Rev.~Lett.~56 (1986), pp.~2280--2283.

\bibitem[8]{vkb91}
A.~Vallat, S.E.~Korshunov, and H.~Beck, {\em XY model on a Sierpinski gasket}, Phys.~Rev.~B 43 (1991), 
pp.~8482--8486.

\bibitem[9]{stran88}
See, for example, K.J.~Strandburg, {\em Two-dimensional melting}, Rev.~Mod.~Phys.~60 (1988), pp.~162--207 and 
the references therein.

\bibitem[10]{rbm98}
R.W.~Reid, S.K.~Bose, and B.~Mitrovi\' c, {\em Monte Carlo study of the XY-model on Penrose lattices},  
J.~Phys.: Condens.~Matter 10 (1998), pp.~2301--2321.

\bibitem[11]{metro53}
N.C.~ Metropolis, A.W.~Rosenbluth, M.N.~Rosenbluth, A.H.~Teller, and E.~Teller, {\em Equation of State 
Calculations by Fast Computing Machines}, J.~Chem.~Phys.~21 (1953), pp.~1087--1092.

\bibitem[12]{bh88}
K.~Binder and D.W.~Heerman, {\it{Monte Carlo Simulation in Statistical Physics}}, Springer-Verlag, Berlin, 
1988.

\bibitem[13]{tc79}
J.~Tobocnik and G.V.~Chester, {\em Monte Carlo study of the planar spin model}, Phys.~Rev.~B 20 (1979),
pp.~3761--3769.

\bibitem[14]{bn79}
A.N.~Berker and D.R.~Nelson, {\em Superfluidity and phase separation in helium films}, Phys.~Rev.~B 19 (1979), 
pp.~2488--2503.

\bibitem[15]{es83}
C.~Ebner and D.~Stroud, {\em Superfluid density, penetration depth, and integrated fluctuation conductivity of 
a model granular superconductor}, Phys.~Rev.~B 28 (1983), pp.~5053--5060.

\bibitem[16]{nk77}
D.R.~Nelson and J.M.~Kosterlitz, {\em Universal Jump in the Superfluid Density of Two-Dimensional Superfluids}, Phys.~Rev.~Lett.~39 (1977),pp.~1201--1205.

\bibitem[17]{k74}
J.M.~Kosterlitz, {\em The critical properties of the two-dimensional xy model}, J.~Phys.~C: Solid State Phys.~
7 (1974),pp.~1046--1060.

\markboth{Taylor \& Francis and I.T. Consultant}{Phase Transitions}
\end{thebibliography}
\end{document}